# ELECTRONIC HEALTH RECORD IN THE ERA OF INDUSTRY 4.0: THE FRENCH EXAMPLE

Sarah MANARD, Nicolas VERGOS, Simon TAMAYO and Frédéric FONTANE

*Mines ParisTech – PSL Research University*
*60 boulevard Saint-Michel, 75272 Paris Cedex 06, FRANCE*

**ABSTRACT**

The recent implementation of the Electronic Health Record (EHR) in France is part of a more general process of digitizing information flows, as the world enters the fourth industrial revolution in a phenomenon known as Industry 4.0. Behind this concept lies the concern to allow Man to remain permanently in control of his destiny, despite an increasingly interconnected world (Internet of Things, cooperative robots, augmented reality, etc.). Accordingly, the implementation of EHR must guarantee the respect for the private life of each citizen. From this perspective, healthcare professionals will therefore have to constantly ensure the protection of medical confidentiality during Electronic Data Interchange (EDI). This paper summarises the current state of the use of EHR in France. Based on a survey conducted by the European Commission to assess the deployment of digitalisation in the health sector in EU countries, this article aims to highlight the opportunities and perspectives that Industry 4.0 could bring to the health sector in France. However, this study also identifies a number of limits related to the application of such a system, the first of which is cyber threat or transhumanism. To this end, a SWOT matrix identifies the strengths and weaknesses related to the implementation of the French EHR.

## 1. INTRODUCTION

The recent deployment of the Electronic Health Record represents a major challenge for France: among the 28 members countries and the 2 partner countries of the EU (EU (28+2)), i.e. the 28 members countries plus Iceland and Norway, it is lagging behind in terms of New Information and Communication Technologies (NICT) (Currie and Seddon, 2014). The digital gap has hit hard a country whose geographical variety and culture are ramparts to the optimal development of these NICTs. Thus, in a context of improving information flows through the digitalisation of processes, a phenomenon that is part of a more general evolution from industry and services to Industry 4.0, the health sector must adapt to the new requirements induced. The emergence of new medical applications on our smartphones follows this trend, with the possibility to book an online consultation, retrieve an electronic prescription or even simplify medical appointment bookings via direct access to doctors' schedules and consultation centres. The implementation of the Electronic Health Record (EHR) in France and Europe must therefore find the answers expected in order to guarantee access to reliable and confidential medical data for all. It will also make it possible, with the help of these new e-services, to respond effectively to the new time requirements induced by the increase in time spent at work or in public transport for workers. Similarly, this EHR should enable everyone, patients and healthcare professionals alike, to have an online connection in line with their needs (Allaert and Quantin, 2009).

    The present study will first aim to establish a state of the art of France in terms of the progress of digitisation in healthcare since the launch of the "EHR for all". To do this, the methodology applied will consist in studying both the expectations of the French with regard to the EHR and the different possible uses and derivatives of the latter, such as epidemiological monitoring and/or follow-up (Haas *et al.*, 2017) as well as decision support to establish diagnoses. In a second step, a European benchmark of the hospital health chain will be made, in order to evaluate the relevance of the implementation of such a system in France. Through the identification of patient and prescription drug information flows, medical research and the

diagnosis and care of chronic and/or rare diseases (Finet *et al.*, 2018) can be improved and thus allow significant advances in medical progress. This will soon be possible with the various online medical records set up in European countries, which we will demonstrate in a summary photo at the end of this article. This French impetus towards eHealth, which is part of a European desire, aims to facilitate the exchange of medical data between the "28+2". The opening of borders would thus be generalized to information flows with an "e-systematization" of medical data directories. So far, no official computer system has been imposed by French political authorities, nor has a standardised model for entering computer data. Moreover, the risks associated with such an implementation should not be overlooked. Cyber threat is a reality that every actor in the healthcare chain will have to fight against. The reliability of computer servers, the security of data, or even the respect of medical confidentiality are concerns that will affect us all. Already today, questions are being raised about access rights to EHR with private and/or specific hospitals and institutions that have their own patient monitoring networks (Sabes-figuera, 2013), as well as the definition of need-to-know according to the specific specialities of each health professional.

## 2. EHR: EXPECTATIONS AND OPPORTUNITIES

These hopes and opportunities were identified based on a literature review of some 15 publications from 2010 to 2019.

### 2.1 Update on the expectations of French people following the implementation of the Electronic Health Record

From the beginning of the project, French authorities have ensured that the EHR aims to guarantee better continuity, better quality and better coordination of a patient's healthcare path. So far, the road to good health resembles a real quagmire, as each healthcare facility gets its own computerized medical data software, sometimes even developed internally. France wants to catch up with its European partners: by 2022, the FNHIF (French National Healthcare Insurance Fund) hopes to create 40 million EHRs (Burnel, 2018). Even if the opening of the EHR is optional, the French seem inclined to take the digital step. Between November 6th - the date of its very last launch - and December 13th, 2018, more than 3 million EHRs were opened (Assurance Maladie française, 2018). Indeed, against all popular beliefs, the French EHR has experienced implementation problems and has been relaunched several times (Séroussi and Bouaud, 2018), as shown in Figure 1.

Figure 1. Timeline of the French EHR, data collected from (ASIPS, 2013).

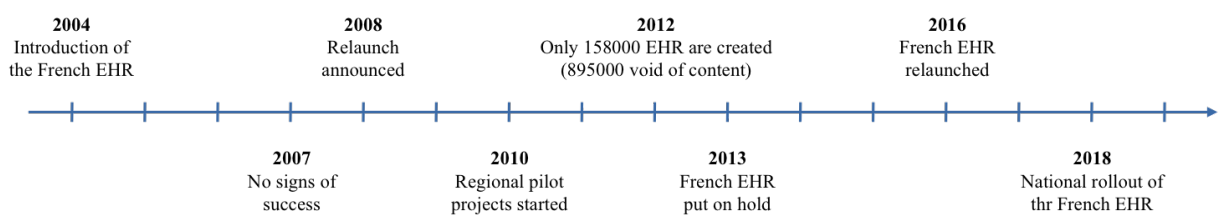

Since its creation, the EHR has been regarded as a data warehouse by healthcare professionals; they can search for information about a patient or consult his healthcare background. All this should ultimately enable them to take better care of their patients. However, in order to guarantee the respect of a patient's private life, it will be possible to withdraw documents that the patient does not wish to divulge to others from this warehouse that a practitioner would have deposited there.

In the field of open data, France already has significant expertise. France has a considerable asset on which it can capitalize and benefit from good practices: since 1999, it has had the largest computer database in the world, the National System of Information between Healthcare Insurance Schemes (SNIIRAM) - managed by the FNHIF. In 2015, 8.9 billion healthcare sheets were managed and anonymized to supply the SNIIRAM warehouse. This database has a storage capacity of 600 terabytes, has 20 billion lines of services available, and manages 150 applications (Bezin *et al.*, 2017). In 2017, a law to modernize the French health system created the National Health Data System (SNDS). The SNDS collects and provides the databases that have existed independently until now:
- the SNIIRAM database containing health insurance data;
- the Information Systems Medicalization Program (IMSP) database containing data from the activities of health care institutions;
- the database Centre for Epidemiology on Medical Causes of Death (CépiDc), managed by NIHMR (National Institute of Health and Medical Research), containing data on causes of death;
- disability-related data from departmental homes for people with disabilities;
- data from "complementary health" (mutual health insurance, for example).

In addition, the FNHIF is now paying ever more attention to its healthcare expenses. This is why it introduced the Public Health Objectives Remuneration (PHOR) in 2011. These are incentive measures for physicians to change their practices to achieve the objectives. They are based on three components: the monitoring of chronic diseases, prevention and the efficiency of prescriptions. ROSP (French remuneration for public health purposes) could thus serve as a lever for healthcare professionals to encourage them to use the EHR. Pharmacists already benefit from payment each time they open an EHR. As local players and trusted third parties, they are called upon to play a major role in ensuring the successful deployment of the EHR. They are paid 1 euro each time they open an HER for a patient (Dormont, 2013).

Faced with the considerable challenge of securing the data entered on the EHR, the French Ministry of Health and Solidarity has chosen a highly secure server, whose name and location are kept confidential for security reasons. Physicians must enter their professional code to log in. If this is their first connection, an alert is immediately sent to the patient by SMS or e-mail. Until authorized, physicians cannot store any data in the EHR system. Currently, they are implementing medical data on patients' DMPs via software made available by the French government. These are linked to secure platforms that allow the official database to be incremented.

Thus, we have gathered the various previous information in Table 1, allowing us to compare the current state of the French patients' health journey with their expectations of the EHR.

Table 1. Expectations of the French regarding the implementation of the EHR.

| State of play | What the French expect |
|---|---|
| • Complicated and laborious healthcare journey<br>• Each professional has his own patient follow-up file | • Better continuity, quality and coordination |
| • Medical data warehouse to which the patient does not have access or access rights | • Transparency of data to the patient<br>• Respect for his privacy and his choices regarding the content of the EHR |
| • The world's largest warehouse of certified and computerized medical data | • Quick and concrete implementation of a reliable and efficient EHR in order to obtain better follow-up/management |

## 2.2 Notable opportunities

The implementation of the EHR represents considerable progress in the French people's healthcare journey. Indeed, the new system adapts to the lifestyle of a majority of the population, which is increasingly mobile. Between 2010 and 2011, 11.9% of 15-59 years olds changed homes, representing nearly 4.3 million mobile phones (FAR & MG, 2015). In addition, the family model is also changing, as more and more stepfamilies are formed: in 2011, 1.5 million children were living in stepfamilies, or 1 in 10 children (Breuil-Genier *et al.*, 2011).

In this context of profound change within our society, the EHR makes it possible to guarantee a follow-up of the patient's medical data anywhere in France or even abroad, thanks to the exchange of standardized medical data.

From birth till death, the patient will have a personal EHR. At each stage of his life, his EHR will offer him a number of advantages.

From birth, the results of tests for genetic diseases carried out on infants may be entered in this EHR. For example, the Guthrie test, performed at birth and used to detect cystic fibrosis, may be impacted in the EHR (Corne and Faure, 2018). A few years later, this will allow a doctor to confirm or invalidate a diagnosis if a patient is in doubt.

In addition, many children undergo minor operations during the first years of their lives: adenoids, appendicitis, etc. These surgical procedures included in the EHR will thus make it possible to avoid possible unnecessary, costly and even dangerous operations (Foucaud *et al.*, 1998).

When the patient is an adult, the EHR will act as a sentinel for healthcare professionals. Major seasonal epidemics such as the flu or gastroenteritis epidemic can be better contained, because the EHR will provide a clear and precise picture of the vaccinations the patient has received. Finally, this system may help to reduce vaccine hesitation, which is still very present in our society (Loubet and Launay, 2017). The EHR will also provide the most accurate experience feedback possible on previous epidemics (location, age, socio-professional category of patients, etc.). This will allow healthcare professionals to apply the right strategy to try to effectively combat these highly contagious diseases.

More generally, eHealth applications will facilitate cooperation between European doctors. The exchange of computer data between healthcare professionals will promote experience feedback and the definition of good practices, thus contributing to the advancement of medicine. In some specific cases, eHealth applications may contribute to the fight against bioterrorism (Hewson *et al.*, 2013).

To go further: based on the model of preventive maintenance in the industrial world, doctors could ultimately perform preventive medical procedures on patients, thanks to eHealth applications. Based on the experience of their British colleagues, French researchers could develop a computer program to detect women at risk of developing preeclampsia, a vascular-placental disease, during their pregnancy (de Moreuil *et al.*, 2018). This first step towards personalized medicine could be generalized to the fight against the two greatest causes of morbidity in France: cardiovascular diseases and cancer (Le Dantec, Chevailler and Renaudineau, 2015). Better still, the use of data obtained during clinical trials could enable health professionals to model therapeutic treatment, thus promoting better patient management (Le Fèvre, Poty and Noël, 2018). Finally, eHealth should contribute to increasing our life expectancy, thanks to the development of both preventive and predictive medicine, which will be increasingly personalized.

## 3. IMPLEMENTATION OF EHEALTH IN EUROPE

Since the launch of the project to create a European EHR in 2004, progress in this field has differed widely from one country to another. Within the EU (28+2), four groups have recently been identified based on their degree of maturity in establishing this project (Currie and Seddon, 2014). Leaders in eHealth in Europe are Finland, Denmark, the Netherlands, Sweden and the United Kingdom. France is currently at a neutral stage where the desire to create a national eHealth data network has been firmly imposed by the government (Brouard, 2015). However, the associated services and tools are only at the beginning of their definitive implementation and generalization in the professional medical environment.

At the twilight of the European Horizon 2020 project, the focus is on the results of the research funded during these six years, among others, to benefit eHealth innovation. Thus, group work (workshops) was able to highlight many disparities within the EU concerning the state of progress of eHealth policy, the heterogeneity of innovations in technology but also in legislation. The lack of a European legislative framework is at stake (Hägglund *et al.*, 2016).

### 3.1 Core principles

In 2003, the ISHARE I project focused on the possibility of exchanging medical data between European countries and thus enabling healthcare professionals to share and consult the diagnoses and observations of their European colleagues (Ochoa *et al.*, 2003). For the implementation of this project, 13 countries were selected as well as national administrative regions identified as representatives of their country's healthcare system and benefiting from local policies conducive to the development of these exchanges. They were subjected to a questionnaire developed as part of the European Community Health Indicators Research (ECHI) to determine the most appropriate national level for exchanging medical data that is valid and usable by other regions.

The final result shows that the exchange of medical data between regions of the same country is possible, allowing the implementation of a compatible database. This project has made it possible to set the foundations for the launch of the European shared medical file by showing that national exchanges of medical data can be systematised and thus make it possible to envisage exchanges abroad. However, particular attention should be paid to the political and administrative context of the participating countries. At the time of the study, Greece had to be excluded because it had just set up a new system for processing medical data within its administration in 2001. Similarly, Finland could not be included in the project due to the specific administrative characteristics of each of its regions (Ochoa et al., 2003).

Thus, the development of a shared European medical record requires first and foremost a national medical data processing unit before it can hope to share it across borders.

Since 2008, with the epSOS ("Smart Open Services for European Patients") project and the creation of the OpenNCP exchange platform, the problems of interoperability and security of medical data exchanges have been partially solved in compliance with national and European laws and policies (Staffa et al., 2018). The KONFIDO project, funded by the EU, was recently launched in order to successfully strengthen the trust and security of medical data exchanges between the different EU Member States. It uses advanced technologies tested in a realistic scenario by a few voluntary countries, such as Italy, and studies the results on the viability and security of the data exchanged (Nalin et al., 2019). Thus, the EU is moving forward to ensure the establishment by 2020 of reliable and secure exchanges of medical data between its member countries, but the road ahead is still full of obstacles.

## 3.2 French progress compared to European countries

In 2012, an update on the progress of the development of eHealth and its associated services at European level EU (28+2) was carried out in order to obtain concrete figures. To this end, 1717 heads of acute care hospitals were interviewed and approximately 50 of them completed a targeted questionnaire on the availability and utilization of specific eHealth tools within their institution (Sabes-figuera, 2013).

Each hospital was evaluated individually on 13 criteria divided into 4 themes to evaluate the level of development of eHealth:
- Infrastructure: external and internal network connections.
- Applications: computer systems for recording patient medical data and examinations within the hospital (PACS) or prescriptions integrated into the patient's computer file.
- Integration: exchange of medical data between hospitals and other medical entities such as analytical laboratories or radiology centers.
- Security: structured frameworks and rules about data access and a system to recover data availability in the event of a technical incident.

We focused the 7 criteria that were most relevant to our benchmark (cf. Table 2). Thus, we only considered institutions with a network communicating with the outside world, ignoring those which did not even have an internal connection between their services. Similarly, only applications that allowed medical data to be communicated with entities outside these hospitals were selected, as well as the frequency of exchanges practiced by the institutions concerned and the reliability maturity of their server. In 2010, this survey had already been conducted, which allows us to assess the evolution of each country between 2010 and 2012.

In Table 2, we have therefore compared the evolution of the 7 criteria used for France and the 2 groups: "Leaders" (Finland, Netherlands, United Kingdom and Denmark) and "Laggards" (Cyprus, Romania, Poland, Greece, Malta, Lithuania, Latvian), in the European reference framework EU (28+2) (Currie and Seddon, 2014). Thus, we note that the changes within French hospitals are not remarkable, and are even declining for most criteria. Efforts are visible when it comes to the exchange between hospitals and external medical entities but lag behind in terms of network infrastructure.

Table 2. Indicators of the level of development of eHealth in France compared to that of the 4 leaders and 7 European "laggards" in the benchmark EU (28+2), data collected from (Sabes-figuera, 2013).

| Domain | eHealth indicators | 7 laggards vs EU(28+2) | France vs. EU(28+2) | 4 leaders vs EU(28+2) |
|---|---|---|---|---|
| Infrastructure | Externally connected | -12,70% | -3% | 3,25% |
| Applications | PACS usage | -12,14% | -25% | 21,50% |
| | ePrescribing | -12,70% | -6% | 36% |
| | Integrated system for eReferral | -4% | -8% | 41% |
| Integration | Exchange of clinical care information with external providers | -14,86% | 4% | 34% |
| Security | Clear and structured rules on access to clinical data | -8,86% | -4% | 11% |
| | EAS for disaster recovery in less than 24 hours | -8,57% | -5% | 19,25% |

Two results attract our attention: the important delta between the "laggards" and "leaders" for the criterion "integrated system for eReferral" and the strong French decline concerning the recording of medical data and examinations in computerized version within hospitals. The first delta can be explained by the fact that the countries considered as "laggards" in the field of eHealth development also correspond to the countries considered as being "less advanced" in terms of economic development in the EU reference framework. Thus, the acquisition of developed and expensive systems to integrate computerized medical data may not be a priority compared to investment in better quality healthcare equipment. As for the sharp decline in the recording of medical data in a computer system in France compared to the EU (28+2), this can be explained by the current political context, which is slow to define the limits of the implementation of an EHR in France. Once these limits are clearly set and the standardization of the EHR decided, the support of French healthcare professionals in the implementation of medical databases will probably be stronger.

In addition, France is just below European average, particularly when it comes to digital development in eHealth and data security and network access. When we compare France's progress with that of the 4 European leaders but also that of the 7 laggards, we see that it is far behind the precursors. In the same way, we note that it is closer to laggards than to leaders in this field.

Figure 2. Average development and progress of eHealth in French and European hospitals within the EU (28+2), data collected from (Sabes-figuera, 2013).

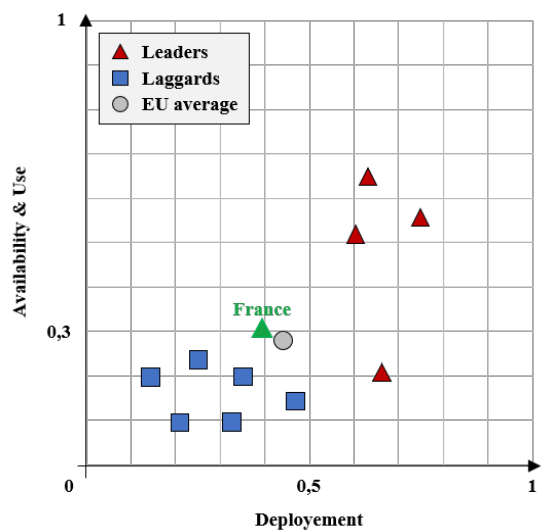

These same indicators were used to assess the development of eHealth, its availability and use by hospital departments, allowing a comparison at national level but also with the EU (28+2) average. It shows that the French average is almost at the same level as the European EU (28+2) average with a score of 36% compared to 30% for Europe (Sabes-figuera, 2013). This study also reveals a great disparity between different French institutions, as some are at almost stage zero in the development of eHealth and its services while others are at the same level as hospitals in the leading countries. This can be explained by the strong regional disparities in France. Outside the major cities and their respective spheres of influence, small towns are medically disadvantaged with the desertification of healthcare professionals who do not wish to settle there (Alfano *et al.*, 2018).

In addition, we see in Figure 2 that the hospital averages (consistent with eHealth development criteria, availability and use within hospitals) of the "leaders" and "laggards" respect the ranking of these groups. The number of hospitals that could be evaluated for this comparison is the highest in France, compared to 10% for some of the "leading" countries (Sabes-figuera, 2013). This can be explained by the desire to create a French national unity before developing at European level. However, the "leading" hospitals remain well above the European average despite the few hospitals that could be selected for this graph. As for the "laggard" hospitals, although the number of hospitals selected for the latter is similar to that selected for the "leaders", their level of eHealth development remains low and well below the European average EU (28+2).

Thus, we can conclude that the level of development of eHealth within the EU (28+2) is directly related to the economic situation of the country (quality of life, welfare state, etc.) but also to the willingness of the government to enact a policy of open borders for medical data.

It is also important to note that many of the expected benefits of e-health in chronic disease prevention and care come from improved primary care, not necessarily hospital care. With the ageing of the population, these non-communicable diseases represent the financial burden of European health care and considerable efforts have been made in this area. However, our present study does not aim to reference medical and treatment advances and their impacts on the European health system (medical and financial) but rather to evaluate the development of eHealth in Europe that enables these new benefits.

## 4. DISCUSSION

Focusing on France, we have constructed a SWOT matrix of the French EHR (cf. Figure 3) which highlights the complexity of its implementation. We have identified only 3 strengths against 5 weaknesses which prevent the realization of this project in France. On the one hand, the French government's welfare state provides a sound and solid basis for the EHR with a robust and efficient health system. On the other hand, the French protectionist culture hinders the implementation of this new IT. Undeniably, the French people value privacy and consider EHR to be a significant intrusion that is not yet sufficiently standardized nor secured.

However, the opportunities provided by the implementation of the EHR are attractive and address many of the problems faced by French citizens (cf. Table 1). Among other things, medical desertification, the ageing population and the high mobility of young workers are finding solutions with EHR-related applications. The arrival on the market of teleconsultations or even connected objects that implement vital data from the French meet the needs of the French but also those of healthcare professionals with easy access to data and better follow-up/management of patients.

In the same way, the data implemented in the databases will be accessible to all European healthcare professionals and will thus be able to prevent pandemics with epidemiological monitoring thanks to the contagion case directory. This will strengthen the links between practitioners and researchers from different countries and build trust in each other's skills. Undeniable progress can be envisaged with the exchange of research results on rare/orphan diseases and innovative treatments recently discovered.

However, as in any new project implemented, some drift is to be expected. Some of them have already been identified, such as cyber threats with the hacking and falsification of stored data, the mercenarism of health professionals with the pursuit of profit at the expense of the patient's wellbeing or telemedicine which destroys human relations between doctors and patients. The latter is to be seriously considered as a risk of gradually seeing a reduction in the medical workforce and patients treated based on the potential gain of the practitioner and not on the proven effectiveness of the treatment.

Before any European implementation, it is imperative to capitalize on French know-how in standardized, robust and certified databases in order to be credible to decision makers. This is accompanied by a strong commitment on behalf of political leaders (cf. Figure 4). As such, the NSDS is a proven prerequisite that can serve as a solid starting point for a rapid and relevant dissemination of the EHR in France.

Figure 3. SWOT matrix of the French EHR.

|  | POSITIVE | NEGATIVE |
|---|---|---|
| **INTERNAL** | **Strenghs**<br>• Largest **certified medical database** in the world<br>• Efficient and well meshed medical network in French cities<br>• Good **dialogue with the European Union** | **Weaknesses**<br>• **Late communication** of the government<br>• **Cultural barrier**<br>• **The French doubt** the reliability of the DMP (data security)<br>• **Health professionals do not support** the concept<br>• **No standardization** of medical I.S. |
| **EXTERNAL** | **Opportunities**<br>• Better **follow-up/patient care**<br>• Pooling of **research on diseases**<br>• Mitigation of **medical desertification**<br>• **Adaptability** of the health system to the contemporary lifestyle<br>• **Exchanges** of best practices<br>• Extended **Epidemiological monitoring**<br>• Implementation of **connected objects** (consultations, prescriptions, patient history)<br>• **Travel** peace of mind in Europe | **Threats**<br>• **Cyber threat**<br>• "Millefeuilles" of application with risk of loss of data standardization<br>• **Transhumanism**<br>• **Mercenarism** among health professionals<br>• **Reduction of medical manpower**/loss of skills with exclusive teleconsultation<br>• European **interference** in our health system |

In order to summarize the study presented in this article, Figure 4 proposes a synthesis of the different fields of application of EHR: it is at the heart of a system of exchange and consultation of computerized medical data of which the patient remains the owner.

Two types of databases feed the EHR: one concerns only the patient with his medical history; the other contains more general information such as decisions to market and/or reimburse innovative medicines/treatments, the results of research on serious/orphan diseases or the reporting of new cases of contagion (Ebola, Plague, Cholera, Flu, etc.), which allows for better epidemiological monitoring. As a result, healthcare professionals can consult these databases and have a decision-making tool at their disposal to better care for their patients. Healthcare professionals feed data into both databases, which are then consolidated in the EHR. The links between healthcare professionals and the databases are constrained to limited consultation (but they can perform queries about any patient) in order to comply with privacy regulations. As a matter of fact, patients decide which healthcare professionals can consult their full data. On the other hand, patients have unlimited consultation access to their own information, but cannot consult data from other patients.

Figure 4. Summary of the implementation of the EHR, the prerequisites for its implementation and its applications.

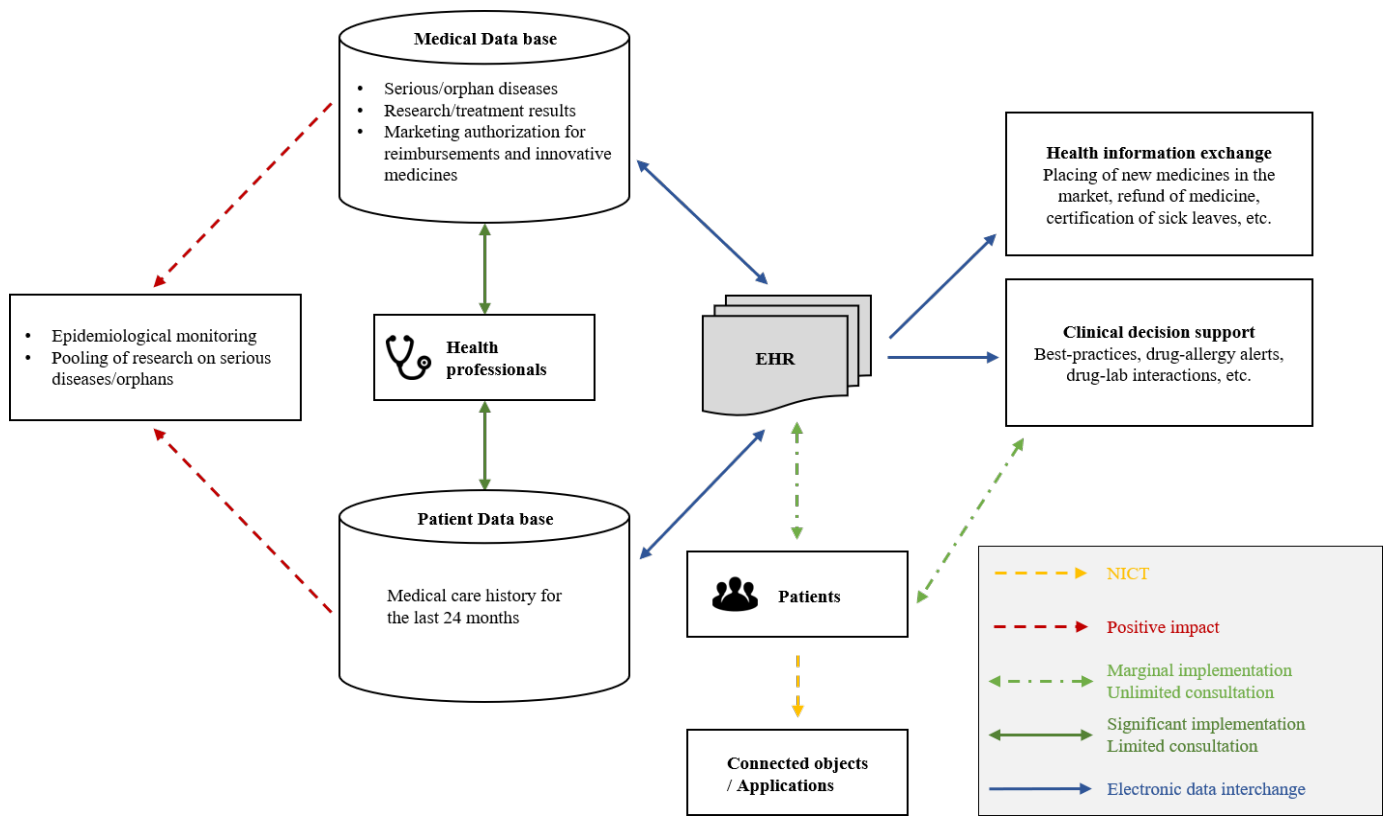

Finally, the Internet of Things completes the field of possibilities in the eHealth domain by collecting vital patient data in real time (connected watches, mobile applications, etc.), or telemedicine (online consultations, ePrescriptions, etc.), which should make it possible to solve the problem of medical deserts and respond to the new lifestyles of the workers (increased mobility, daily commutes, etc.).

It is essential to highlight here that an implementation like the one seen in Figure 4 can only take place if 4 main prerequisites are met:
1. Standardization of national data.
2. Certified and consistent databases.
3. Stable economic environment.
4. Strong willingness of governments to implement EHR.

As was demonstrated in this article, the EHR project is no longer a simple utopia (Cacot, 2016). However, it will be necessary to ensure that the implementation of the EHR does not drag on too long and get lost in a heap of complex variations, otherwise it risks becoming a chimera.

## 5. CONCLUSION

This article proposed an analysis of the fields of application of eHealth in France and Europe, in particular through the implementation of the EHR. The SWOT matrix highlighted a number of opportunities, such as the mitigation of medical deserts, better epidemiological monitoring and the adaptability of the healthcare system to the new contemporary way of life.

France has considerable assets enabling it to implement the EHR under good conditions: a reliable and relatively exhaustive medical database, a high-performance healthcare system envied throughout the world, good engineering skills, innovative start-ups, etc. However, France still seems to lag behind its European partners in terms of eHealth applications and use.

By placing greater trust in its engineers and healthcare professionals, France will be able to deploy its EHR, and more generally eHealth applications, in a consistent and efficient manner.

With the opening of borders and cooperation between European countries, an eHealth exploratory laboratory could be created. This would promote research and development (R&D) of new common and interoperable eHealth applications for all European patients.

In addition, the establishment of a European EHR will also require stable economic contexts in all participating countries and a standardization of the data implemented in this dossier.

The benefits of this new medical tool are numerous and revolutionary. It will be able to provide epidemiological monitoring over a wide area by identifying cases of contagion, pool the results of innovative research on serious and orphan diseases, and help doctors in complex cases to decide on the most appropriate treatment. From the patient's point of view, the first expectations are total transparency with access to all their medical data but also the primary right to possess their data by having the power to decide whether or not to disclose them. In the same vein, the EHR will make it possible to instantly and securely disseminate information on the marketing of innovative medicines but also on their reimbursement by public bodies. In this way, patients will be able to benefit from the latest medical advances for their treatments without being constrained by the restriction of their right to know.

In the age of Industry 4.0, new telemedicine-related applications such as e-consultations via smartphones, online medical prescriptions and connected objects allowing real-time patient tracking are beginning to appear on the market. It is a revolution in a context of high geographical mobility of populations. With these applications, the follow-up of patients at home is also facilitated for healthcare professionals and makes it possible to push back the limits of medical desertification and the ageing of populations constrained in their movements. Thus, patient care would be facilitated by implementing medical data from the EHR to these connected objects and applications.

However, these applications, designed to save time for both healthcare professionals and patients, should not become a simple "cash machine" whose yield would be the only concern of physicians using such tools. Used to the extreme, such applications would involve risks of transhumanist drifts (tracking by chips implanted under the skin, vital organs duplicated by 3D printing, etc.) and could eventually lead to a commodification of the human body. Thus, the patient would become a simple object to be maintained in an optimal state of functioning, with preventive operations/treatments, without any ethical consideration. He would be tracked throughout his life. In such a system, one could then ask oneself: what are a patient's rights to privacy and possible treatment choices? This would raise the question of access to appropriate care based on reliable and well-considered diagnoses.